# Growth of AlGaN under the conditions of significant gallium evaporation: phase separation and enhanced lateral growth.

I.O. Mayboroda,[a)] A.A. Knizhnik, Yu.V. Grishchenko, I. S. Ezubchenko, Maxim L. Zanaveskin, M. Yu. Presniakov, B.V. Potapkin, V.A. Ilyin.

*National Research Centre "Kurchatov institute", Moscow, 123182, Russia*

[a)] Author to whom correspondence should be addressed.  Electronic mail: mrlbr@mail.ru

**Abstract**

Growth kinetics of AlGaN in NH3 MBE under significant Ga desorption was studied. It was found that the addition of gallium stimulates 2D growth and provides better morphology of films compared to pure AlN. The effect was experimentally observed at up to 98% desorption of the impinging gallium. We found that, under the conditions of significant thermal desorption, larger amounts of gallium were retained at lateral boundaries of 3D surface features than at flat terraces because of the higher binding energy of Ga atoms at specific surface defects. The selective accumulation of gallium resulted in an increase in lateral growth component through the formation of the Ga-enriched AlGaN phase at boundaries of 3D surface features. We studied temperature dependence of AlGaN growth rate and developed a kinetic model analytically describing this dependence. As the model was in good agreement with the experimental data, we used it to estimate the increase in the binding energy of Ga atoms at surface defects compared to terrace surface cites using data on the Ga content in different AlGaN phases. We also applied first-principles calculations to the thermodynamic analysis of stable configurations on the AlN surface and then used these surface configurations to compare the binding energy of Ga atom at terraces and steps. Both first-principles calculations and analytical estimations of the experimental results gave similar values of difference in binding energies; this value is 0.3 eV.

**I. Introduction**

AlN, GaN, InN, and their alloys have found wide application in micro- and optoelectronics because of unique properties of these materials. Nitride semiconductors are used in microelectronics to fabricate high performance ultrahigh-frequency (UHF) devices based on high-electron-mobility transistors (HEMT). Despite the spectacular progress of past two decades [1], poor morphological and structural properties of heterostructures remain a limiting factor for nitride-based UHF devices because of the use of lattice mismatched substrate materials, such as Si, SiC, and $Al_2O_3$ [2].

     Nitrides form a continuous series of ternary alloys AlGaN, InAlN, and InGaN. The properties of these alloys, such as band gap width, spontaneous and piezoelectric polarization constants, and lattice constants continuously vary with composition [3]. The possibility of controlling parameters through alloy composition is vastly used in the fabrication of active and functional layers of electronic nitride devices with the required characteristics.

Some problems of ternary nitride film growth arise from the difference in the stability of Al, Ga, and In bonds with nitrogen. AlN, GaN, and InN are characterized by different equilibrium nitrogen pressures over their surfaces [4] and have different thermal stabilities under the same nitrogen pressure. Because of a difference in the stability of chemical bonds between Al, Ga, and In and nitrogen, AlN, GaN and InN start to decompose at different temperatures. Under vacuum, these binary films are known decompose through the congruent evaporation of metal and nitrogen atoms. Ternary alloy decomposition is more complex. Ternary nitride films are grown in one of the following three modes: (i) at sufficiently low temperatures both metal components are stable within a crystal and film decomposition is not observed; (ii) at extremely elevated temperatures, less stable metal atoms (indium in InGaN and gallium in AlGaN) are almost fully desorbed with the formation of a binary film with impurities of less stable metal atoms; (iii) at intermediate temperatures, both incorporation and desorption of less stable metal atoms occur and result in the dependence of film composition on the growth temperature. The temperature boundaries of these modes depend on the ratio between nitrogen and metal components.

At the moment, low temperature growth is a conventional regime, as it allows simple and precise control of the film composition. Unfortunately, low temperature growth yields films with poorer morphology and crystal quality because of the low surface mobility of adatoms. Meanwhile, growth at extremely elevated temperatures had also found practical application. Despite being incorporated at impurity level, less stable metal atoms affect growth kinetics and promote the formation of films with smoother surfaces, relaxed strains, and lower defect densities. The addition of indium was shown to improve the quality of GaN films grown by metal-organic chemical vapor deposition (MOCVD) [5], plasma-assisted molecular beam epitaxy (PAMBE)[6], and ammonia-based molecular beam epitaxy (NH3 MBE) [7]. Ga was also found to improve the morphology and crystal quality of AlN films grown by MOCVD [8]. The authors of the listed papers associated the action of In and Ga with the surfactant effect.

In the epitaxy, surfactant is a substance that has the following features: it covers the surface of a growing film, affects growth kinetics, but incorporates weakly (at a doping level) into the bulk of the forming crystal. The effect of a surfactant is usually explained by an increase in the adatom diffusion lengths. For example, when Sb was used in Si/Ge epitaxy as a surfactant, it formed a solid coverage on the growing surface and reduced diffusion barriers for Si and Ge adatoms [9]. The same mechanism was found for indium-mediated GaN growth by PAMBE under Ga-rich conditions, when indium and gallium formed a metallic layer, providing the direct interaction between the In-surfactant and other adatoms [6]. For ammonia-based $NH_3$ MBE and MOCVD methods, the mechanism of surfactant effect was either attributed to the enhanced mobility of the adatoms [5, 7] or was not discussed at all [8]. Nevertheless, surfactant-mediated growth by $NH_3$ MBE and MOCVD [5, 7, 8] took place under strong ammonia excess (N-rich) conditions, which excluded the formation of a solid metallic coverage. According to theoretical estimates for GaN growth by $NH_3$ MBE under strong $NH_3$ excess, Ga adatoms occupy less than one percent of surface sites [10]. Therefore, the presented explanations of the surfactant effect of gallium and indium are contradictory for the ammonia based growth techniques (MOCVD and NH3 MBE).

The intermediate range of growth parameters between complete incorporation and full evaporation of less stable metal atoms (Ga in AlGaN and In in InGaN) is poorly used and studied. A common trend of a decrease of AlGaN growth rate with temperature was shown for MOCVD by *Keller et al* [11]. A more detailed investigation was provided recently for MOCVD grown AlGaN films with a mole fraction of GaN over 60% under the thermal desorption of Ga [12]. *Skierbiszewski et al.* succeeded in the development of a kinetic model for InGaN growth under N-limited conditions [13]. Meanwhile, there is a lack of the data on the morphology and growth of AlGaN films with high AlN content under strong Ga desorption.

The common issue of growth of ternary nitride films is the separation of phases of different compositions. Phase separation in ternary nitride alloys appears to be a complex phenomenon, affected by surface morphology and strain inhomogeneity. It is useful in case of self-organized quantum dots [14], but is considered as a negative effect when homogeneous films are required [15].

In this work, we studied the kinetics of the epitaxial growth of AlGaN films in a temperature range between full evaporation and complete incorporation of gallium. Experimental data on a comparison of AlGaN and AlN film series grown under similar conditions within the temperature range of significant gallium evaporation are presented in section II. The effect of Ga on growth kinetics and film morphology is demonstrated. The formation of an AlGaN phase with a higher Ga content in the areas of developed surface is shown. Section III presents the results on an experimental study of the dependence of AlGaN growth rate on growth temperature and introduces a kinetic model analytically describing this dependence. The model is then used to explain the formation of AlGaN phases with higher Ga contents though the accumulation of gallium at the boundaries of 3D surface features because of the dependence of Ga binding energy on local surface morphology. Local phase separation through the selective accumulation of gallium in the vicinity of surface defects is associated with the enhancement of lateral growth. Section IV describes methods and results of first-principles calculations of Ga atom interaction with morphological defects on the AlN (0001) surface. For this goal, we first developed an ab initio thermodynamic model to examine stable configurations on this surface and then used these surface configurations to compare binding energies of Ga atoms at terraces and steps. Section V presents a general discussion of the results and our considerations of the relations between phase separation in AlGaN, enhancement of lateral growth, and surfactant effect. Section VI contains brief conclusions.

**II. AlN and AlGaN growth: analysis and comparison**

The AlN and AlGaN films were grown on 50 mm diameter (0001) oriented sapphire substrates with a 0.2 degree miscut toward the m-plane. An optical pyrometer calibrated to the melting point of Al (660°C) was used to measure substrate temperature. Prior to growth, the substrates were annealed for 1 hour at 850°C under high vacuum and nitrided at the same temperature within 15 min in a 30 sccm $NH_3$ flux. Growth rates were obtained from oscillation curves of surface optical reflectivity. Al and Ga fluxes are presented in units of growth rate, nm/h [16]. Ammonia, Al, and Ga fluxes were 200 sccm, 200 nm/h, and 270 nm/h, respectively. The thickness of the deposited films was 300 nm. Reflective high-energy electron diffraction (RHEED) and atomic force microscopy (AFM) were used to study surface morphology. Surface reconstructions observed by RHEED showed that all of the grown films were metal-polar [17]. The stoichiometry of AlGaN films was analyzed by X-ray diffraction (XRD) [18]. Samples for transmission electron microscopy (TEM) were prepared by mechanical polishing followed by Ar-ion milling. All TEM samples were cut normally to the substrate (0001) plane. The local chemical composition of AlGaN was examined by energy dispersive X-ray spectroscopy (EDXS) using an EDXS-module of TEM system.

The RHEED patterns for AlN and AlGaN films evolved similarly. Growth started with the formation of small nucleation islands, producing spotty RHEED patterns [Fig. 1(b)]. The further growth showed a gradual transition to streaky diffraction patterns, indicating the formation of smooth surfaces [Fig. 1(c)]. After some time, a full transition to streaky patterns took place. This transition was related to a change in growth mode from three-dimensional to two-dimensional (3D–2D transition) [19, 20]. Full 3D–2D transition times for different AlN and AlGaN films are presented in table I. One can see that 3D–2D transition took significantly less time for AlGaN films. Growth rates were also measured. The measured AlN growth rate for all AlN samples was 200 nm/h. The AlGaN growth rates decreased at higher deposition temperatures, as was expected. Using the measured growth rates and 3D–2D transition times, we calculated 3D layer thicknesses, which are presented in Table I. For AlGaN films, the GaN mole fraction decreased with temperature (Tab. I).

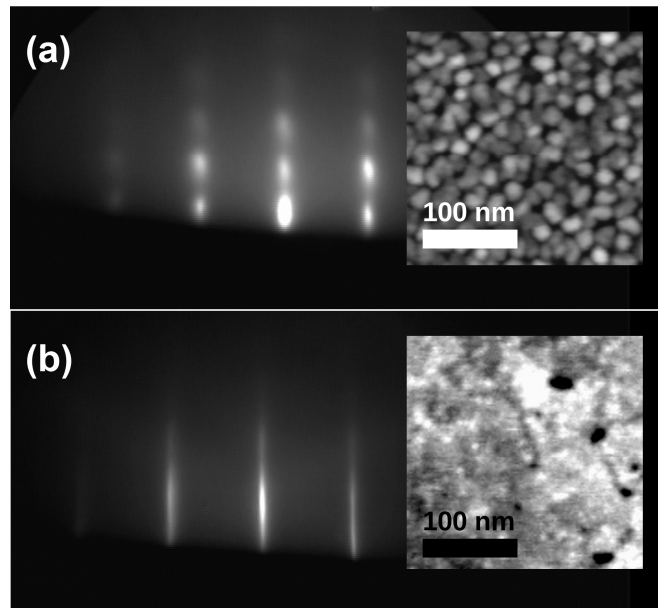

FIG. 1. RHEED images for (a) 3D and (b) 2D growth stages, respectively. Insets show AFM images of (a) 3D islanded surface and (b) uniform smooth surface, corresponding to RHEED patterns.

AlN films [Fig 2(a)] had terrace-stepped morphology with the mean terrace length and step height 0.5 μm and 2 nm, respectively. The ratio of step height to terrace length corresponds to a substrate miscut angle of 0.2 degree, which indicates that the terraces are parallel to the (0001) plane of the substrate. Hillocks of the average diameter 200 nm are also seen. AFM images of higher resolution showed that the surfaces of all samples had pits whose densities exceed $10^9$ cm$^{-2}$ [Fig. 2(b)]. TEM analysis showed that these pits originated from threading pores, penetrating through the bulk of the AlN film [Fig. 2(c)].

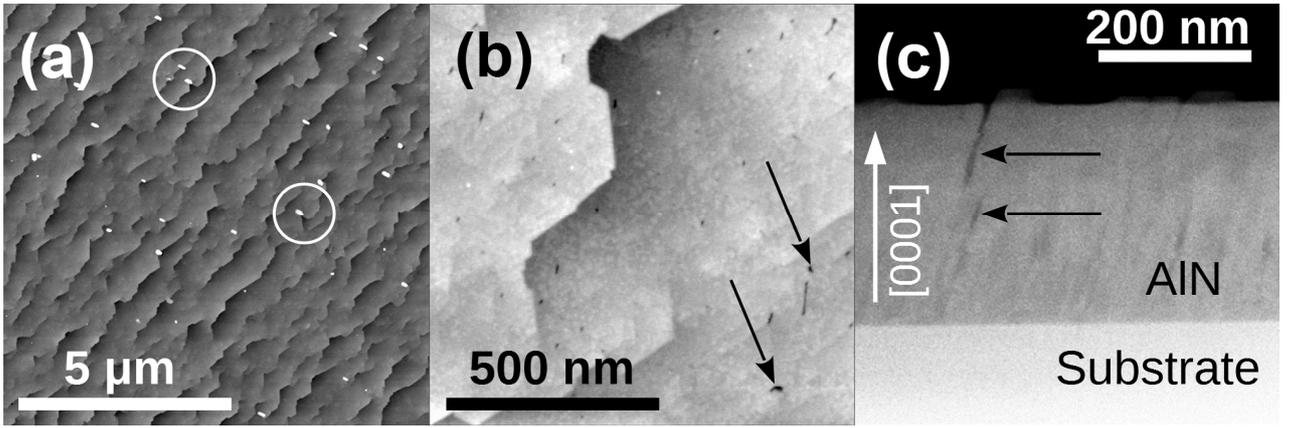

FIG.2. AlN films morphology: (a), (b) AFM images of AlN surface for the film grown at 1050°C under ammonia flux of 200 sccm; (c) TEM image of a AlN film. Circles and arrows show hillocks and threading pores, respectively.

AlGaN films had terrace-stepped surface with average terrace length and step height increased up to 1 µm and 4 nm, respectively. Terraces were atomically flat and exhibited no pits. The average hillock diameter *increased* to 400 nm.

TABLE I. Characteristics of AlGaN films

| Characteristic | AlGaN | | | AlN | | |
|---|---|---|---|---|---|---|
| Growth temperature, °C | 1020 | 1050 | 1100 | 1020 | 1050 | 1100 |
| GaN mole fraction in AlGaN (XRD) | 0.39 | 0.18 | 0.02 | - | - | - |
| Full 3D–2D transition time (RHEED), min | 6 | 6 | 6 | 12 | 16 | 15 |
| AlGaN growth rate (interferometry), nm/h | 330 | 245 | 205 | 200 | 200 | 200 |
| 3D layer thickness, nm | 33 | 25 | 20 | 40 | 53 | 50 |

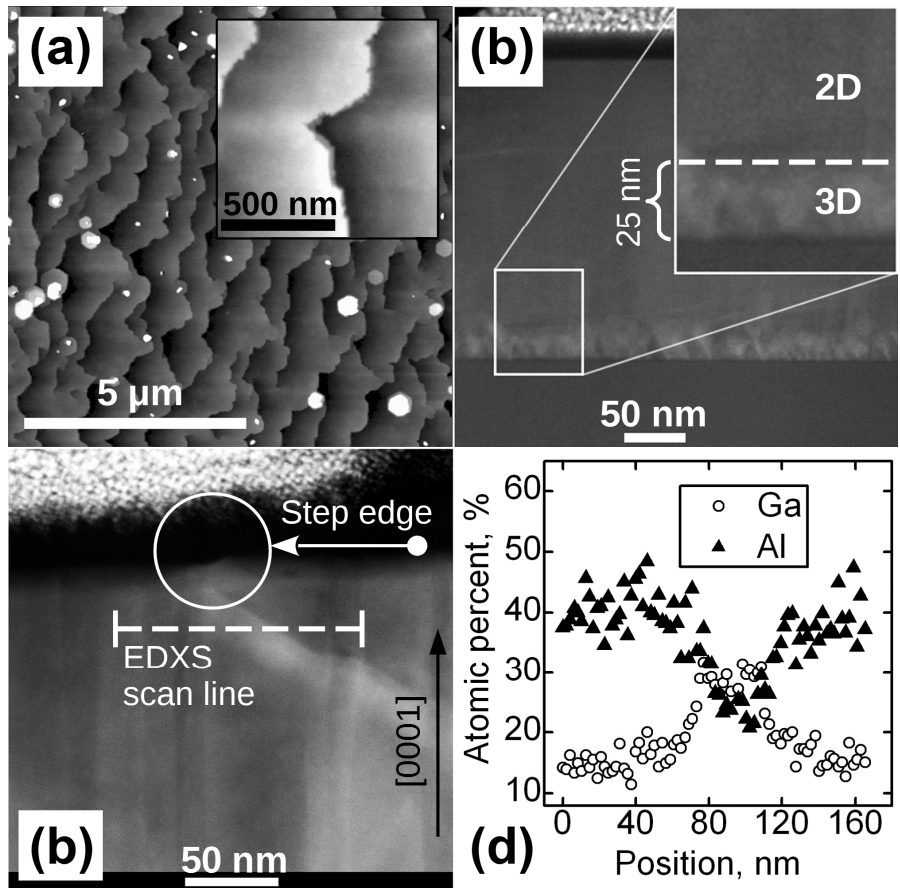

FIG. 3. Experimental data for a AlGaN film grown at 1050°C: (a) AFM images; (b),(c) dark field TEM images of the sample cut; (d) distribution of Al and Ga atomic concentrations along the EDXS scan line, shown in (c). Inset in (b) shows an enlarged area, where a 3D–2D transition is seen at a AlGaN film thickness of 25 nm.

The AlGaN samples for TEM were cut from the AlGaN film grown at 1050°C normally to both the (0001) plane and terrace edges [Fig. 3(b)]. The AlGaN film had no threading pores. The nucleation layer shows *remarkable contrast* in Fig. 3(b). A similar contrast is also clearly seen in the areas under step edges. EDSX analysis showed that, in the bright areas under step edges, the GaN mole fraction was close to 0.55, whereas the average GaN mole fraction was 0.18 according to XRD. The GaN mole fraction in the nucleation layer also increased (Fig. 4). Therefore, the contrast observed in TEM images was due to the higher GaN mole fraction in bright areas. The thickness of the 3D layer according to TEM was 25 nm [Fig. 3(b)], which justified the RHEED-based evaluation (Tab. I).

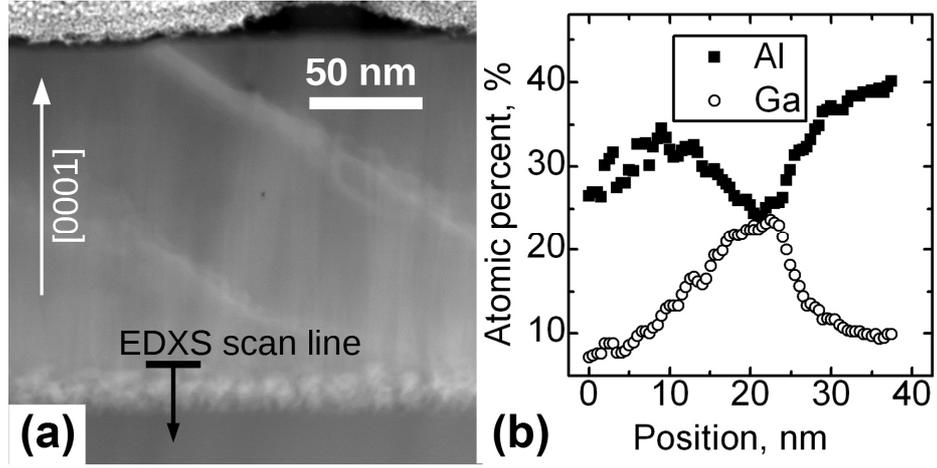

FIG. 4. (a) A dark field TEM image of an AlGaN sample cut and (b) distribution of Al and Ga atomic concentrations along the EDXS scan line.

The application of Ga induced noticeable changes in the film evolution, structure, and surface morphology. The 3D–2D transition at the initial stages of growth took less time under gallium exposure for all AlGaN films. All AlGaN films had atomically flat terraces with larger widths and did not have threading pores. We emphasize that these effects were also observed for AlGaN films grown at 1100°C with a GaN mole fraction of 0.02 (see Table I).

Ga
### III. AlGaN growth rate and composition.

In this section we consider the dependence of AlGaN growth rate and composition on the growth temperature. We introduce an analytical model describing changes in AlGaN growth rate and composition based on the analysis of elementary surface processes. In the end of the section, the developed model is applied to analyze selective gallium accumulation and its effect on film growth. To examine AlGaN growth rates at different temperatures in more detail, additional experiment was carried out. The AlGaN film was grown with $NH_3$, Ga and Al fluxes being 200 sccm, 270 nm/h, and 200 nm/h, respectively. Growth temperature was varied within the range 800–1100°C. Growth rates in the course of deposition were measured using an optical interferometry system. The data obtained are presented by empty square markers in figure 5(a).

*Kinetic model of AlGaN growth.*

AlGaN is a solid alloy of AlN and GaN; therefore was reasonable to consider AlGaN growth rate $V_{AlGaN}$ as a sum of AlN and GaN growth rates, $V_{AlN}$ and $V_{GaN}$, respectively. These growth rates are related as follows:

$$V_{GaN} = V_{AlGaN} - V_{AlN} = xV_{AlGaN}, \qquad (1)$$

where $x$ is the mole fraction of GaN in $Al_{1-x}Ga_xN$. Using Eq.(1), we found that $V_{GaN}$ was 130, 45, and 4 nm/h at growth temperatures of 1020°C, 1050°C, and 1100°C, respectively, whereas $V_{AlN}$ was 200 nm/h, which is equal to the impinging Al flux. Let $p$ be Ga incorporation rate defined as $p = V_{GaN}/I_{Ga}$, where $I_{Ga}$ is an impinging Ga flux. One can see that the change in $V_{AlGaN}$ with temperature was caused by the decrease in $p$, while the Al incorporation rate was unity.

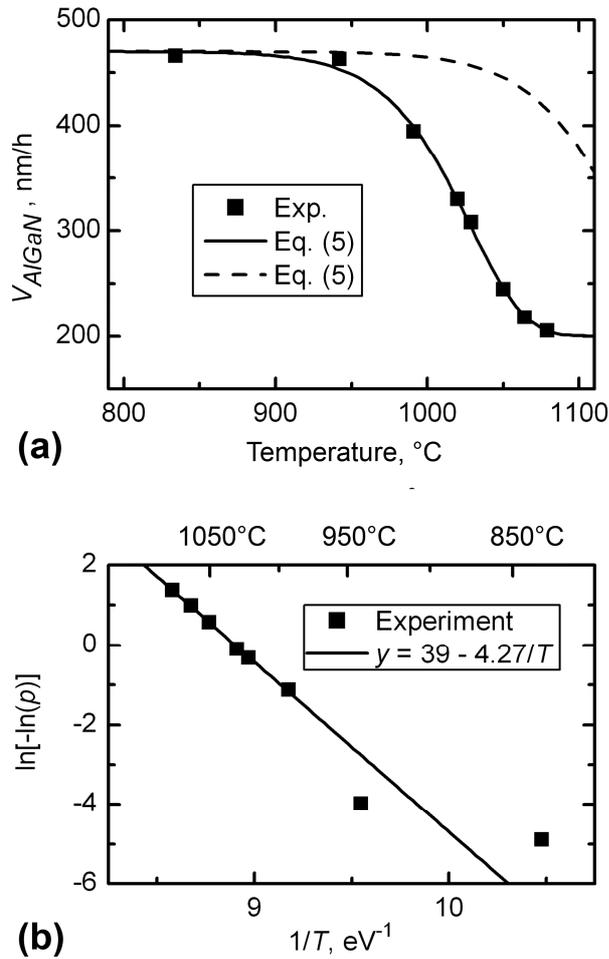

FIG. 5. AlGaN growth rate against growth temperature: (a) experimental points (empty squares) and theoretical curves obtained using Eq. (5) for $E = 4.27$ eV (solid line) and $E = 4.6$ eV (dashed line), respectively; (b) plot of $\ln(-\ln p)$ vs. $1/T$ for the experimental values of $p$ (empty squares) and function $y = 39 - 4.27/T$ (solid line).

We used a kinetic approach [10] to develop a simplified model describing the dependence of $V_{AlGaN}$ on temperature under the conditions of Ga desorption. The development of a strict kinetic growth model is a difficult task, but specific growth conditions allowed us to make several significant simplifications.

The analysis presented above shows that the change in the growth rate and composition of AlGaN films with temperature was due to the desorption of gallium, while the Al incorporation rate was constant and equaled unity. Therefore, we considered the AlGaN film as a stable AlN matrix with an unstable gallium additive. If the Ga atom was placed onto the AlN surface at high temperature without impinging Al and $NH_3$ flux, it would have been desorbed within a finite time interval because of weak Ga—N bonds. Therefore gallium remains stable within the growing crystal only if it is buried under a new AlN layer. Thereby we considered three possible elementary events: (i) chemisorption of Ga atoms, (ii) desorption of Ga atoms, and (iii) blocking of gallium by the forming AlN layer.

Let us consider the following sequence of elementary processes for a single Ga atom in time. At the moment $t = 0$, Ga is chemisorbed in the AlN surface with total binding energy $E$. At the same moment Al and $NH_3$ supply is turned on, and the formation of a new AlN layer is started. Let $\beta(t)$ be the function describing the probability of finding a not desorbed Ga atom. Obviously, $\beta(0) = 1$. Under the assumption that the Polanyi–Wigner equation is valid for the thermally driven desorption of Ga, and if AlN growth is not taken into account, we obtain the following expression for $\beta(t)$:

$$\frac{d\beta}{dt} = -\beta A e^{-E/T}, \qquad (2)$$

where $E$ is average binding energy of Ga on the AlN surface, $T$ is absolute temperature in *energy* units, and $A$ is a pre-exponential factor. If we consider the probability of Ga blocking (covering) by the forming AlN layer, it will be described by a certain monotonically increasing function $f(t)$ of time. If $\tau$ is time instant when a solid AlN layer formed, then $f(0) = 0$, and $f(\tau) = 1$. Accounting for $f(t)$, in Eq. (2), we obtain:

$$\frac{d\beta}{dt} = -\beta A e^{-\frac{E}{T}}[1 - f(t)], \qquad (3)$$

Integrating Eq. (3) from $t = 0$ to $\tau$, we obtain total probability that a single Ga atom is incorporated into the growing AlGaN film:

$$\begin{aligned}\beta(\tau) &= \exp[-\exp(B - E/T)], \\ B &= \ln\left(A \int_0^\tau [1 - f(t)]dt\right).\end{aligned} \qquad (4)$$

Note that $\beta(\tau)$ is actually equal to the Ga incorporation rate $p$, therefore:

$$V_{\text{AlGaN}} = I_{\text{Al}} + pI_{\text{Ga}} = I_{\text{Al}} + I_{\text{Ga}}\exp[-\exp(B - E/T)], \qquad (5)$$

where $I_{Al}$ and $I_{Ga}$ are impinging Al and Ga fluxes. Figure 5(b) shows a plot of the function $y = \ln(-\ln p)$ vs $1/T$ for $p$ calculated from the experimental data using Eq. (5). The experimental points in Fig. 5(b) show an evident linear trend for temperatures above 950°C. Its linear fit gives average binding energy for Ga atoms $E = 4.27$ eV. When $V_{\text{AlGaN}}(T)$ given by Eq. (5) is plotted with the extracted $E$ and $B$ values, it shows good agreement with our experimental data.

It is evident that the local variation of binding energy $E$ must lead to different incorporation rates $p$. When $E$ is increased, the $V_{\text{AlGaN}}$ curve moves toward higher temperatures [Fig. 5(a)], meaning that the higher value of $E$ must result in a higher gallium content. We associate the formation of AlGaN phases of a higher Ga content with a local difference in the Ga binding energy. The experiment showed that the GaN mole fraction in Ga-enriched regions exceeded 55%, corresponding to the incorporation of more than 90% of the impinging gallium atoms. We estimated the increase in energy 0.33 eV for $E$ to be sufficient for the 90% incorporation of gallium.

### Enhancement of lateral growth component due to the selectivity of Ga evaporation.

According to the basics of the terrace–step–kink model (TSK) [21, 22], binding energy of an atom depends on the number of chemical bonds with the crystal. Step edges are supposed to provide more chemical bonds compared to terraces (Fig. 6). Under such consideration, island boundaries, pits, and vacancies will also provide additional chemical bonds. Therefore, under the conditions of Ga evaporation, 3D features must retain more gallium atoms at their boundaries, which results in gallium accumulation. This explains the enhanced incorporation of Ga at the 3D growth stage and along step edges (Fig. 3, 4).

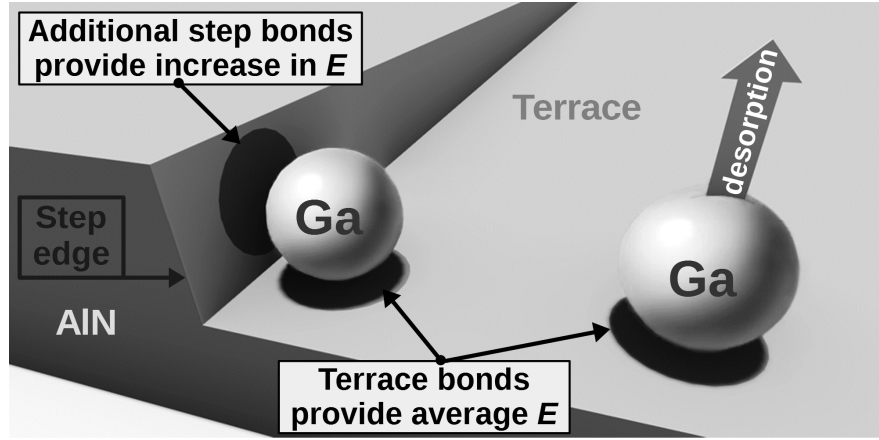

FIG.6. Schematic of interaction between Ga atoms and AlN surface. The step provides additional chemical bonds and increases the total binding energy $E$ of an atom. An atom at the terrace has a lower binding energy and a higher desorption probability.

As Ga atoms are selectively retained by the sidewalls of steps, islands, and pores, they accelerate the widening of surface features and affect lateral step propagation during the film formation. The growth can be considered consisting of two components: vertical growth and lateral growth. The lateral component corresponds to the movement of island boundaries and surface steps along the film plane, whereas vertical component corresponds to a change in film thickness. Let the lateral growth component for AlN and AlGaN be denoted as $L_{AlN}$ and $L_{AlGaN}$, respectively. The ratio $R = L_{AlGaN}/L_{AlN}$ is determined by the total amounts of incorporated atoms for AlGaN and AlN growth:

$$R = L_{AlGaN}/L_{AlN} = V_{AlGaN}/L_{AlN} = 1/(1-x), \qquad (6)$$

The mole fraction of GaN at the step edges and in the nucleation layer of AlGaN film grown at 1050°C reached 0.55; hence we obtain $R = 2.2$. Simple calculations using Eq. (4) show that a 0.33 eV increase in binding energy $E$ is sufficient for the attainment of such value of $R$ [see dashed line in Fig. 5(a)].

Let us consider the effect of an increase of lateral growth rate on film evolution during the growth process. It was mentioned above that the growth of nitride films on mismatched substrates usually starts with the formation of 3D nucleation islands. Under proper conditions islands expand and merge into a solid film, and the growth mode changes from 3D to 2D. The widening rate of islands is determined by the lateral growth component. Thus, enhanced lateral growth will lead to a faster merging of islands and faster 3D–2D transition. As shown above, gallium reduced the period of 3D–2D transition twofold (Tab. I), whereas the average hillock diameter demonstrated a twofold increase from 0.2 to 0.4 μm.

It is important that 3D features are required to promote the formation of the AlGaN phase with a higher Ga content. As soon as 3D islands merged into a solid smooth film, the formation of a Ga-enriched AlGaN phase stopped. This is exactly the effect we observed in our AlGaN films using TEM and EDXS (Fig. 3, 4).

## IV. First-principles calculations.

In the previous section, it was supposed that a chemisorbed Ga atom at a step is more strongly bound to surface compared to a Ga atom at a terrace. To test this assumption, we carried out first-principle calculations of Ga stability at different sites on the AlN (0001) surface.

### *Computational approach*

Calculations of the interaction of Al, Ga, and nitrogen precursors with the AlN(0001) surface were performed using the first-principles method based on the density functional theory (DFT), implemented in the VASP code [23, 24] [. The density functional was based on the generalized gradient approximation (GGA). The wave functions of valence electrons were decomposed using a plane-wave basis with a cutoff energy of 500 eV. The projected augmented wave (PAW) method [25] was used to describe the effect of core electrons. For the Ga atom, 3d electrons were explicitly included into the valence shell. Sampling of the Brillouin zone in the inverse space was done using the (3x3) Monkhorst–Pack mesh for a (2x2) surface supercell on the AlN(0001) surface.

### *Models of AlN(0001) surface*

To investigate the adsorption properties of the AlN(0001) surface, we considered two models of the AlN(0001) surface: a model of flat (0001) surface with a (2x2) supercell [see Fig. 7(a)] and a model of stepped (0001) surface using a (5x3) supercell, which is shown in Fig. 7(b). Each model contained 4 bilayers of Al and N of the thickness 10 Å, nitrogen atoms at the bottom were passivated by hydrogen atoms, and a 10 Å vacuum layer was added. The two lowest bilayers of AlN were fixed in the positions of an ideal crystal.

We considered several surface configurations of Al, Ga, N, and H adatoms and adsorbed $NH_3$, $NH_2$, NH groups. Different adsorption sites on the AlN(0001) surface are shown in Fig 7(a) and include an HCP1 site above the surface Al atom (also called top site), an HCP2 site above the subsurface nitrogen atom and an FCC site above the hole in the wurtzite structure. In addition, there is an adsorption Bridge site in between two surface Al atoms.

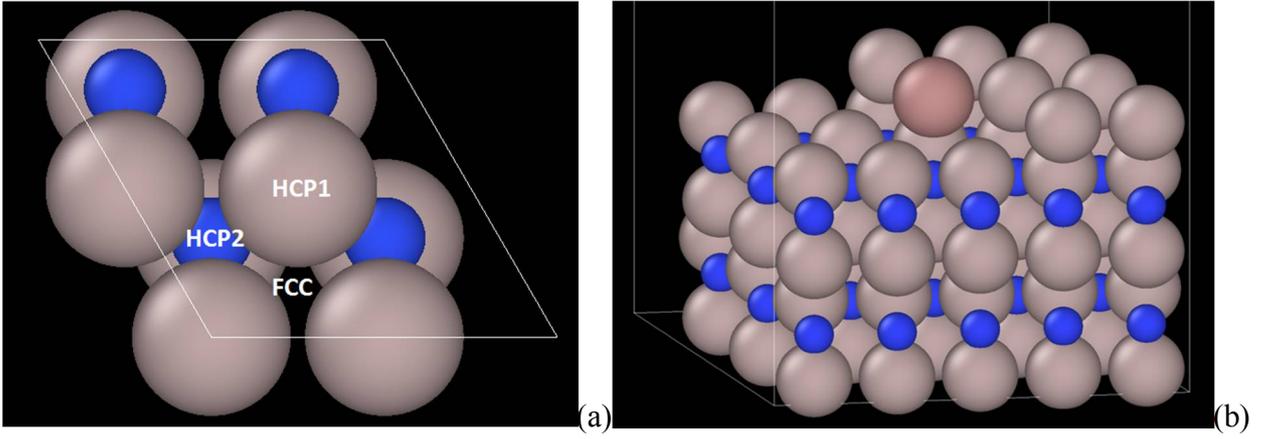

Fig. 7. (a) Model structure of the AlN(0001) based on a (2x2) supercell. (b) Model structure of the stepped AlN(0001) surface based on a (5x3) super cell. Brown balls are Al atoms, blue balls are nitrogen atoms.

*Adsorption properties of the pure AlN(0001)surface and surface diffusion*

The adsorption energies of Al, Ga, and N atoms at different adsorption sites on the pure AlN(0001) surface are given in TABLE II (adsorption energy of nitrogen was calculated relative to the $N_2$ molecule in the gas phase). One can see that metallic Al and Ga atoms have the highest adsorption energy at the HCP2 site, located above the subsurface nitrogen atom. The adsorption energy of the Ga atom is 0.3 eV lower than that of the Al atom, pointing to the easier desorption of Ga from the surface. In contrast to metal adatoms, the N adatom strongly prefers the FCC site.

The diffusion of metal atoms on the AlN(0001) surface can proceed as a jump from an HCP2 site via Bridge and FCC sites to a neighboring HCP2 site. As is seen in TABLE II, the maximum energy along this pathway corresponds to the transition over a FCC site and is 0.65 eV for Al and 0.62 eV for Ga. The diffusion of nitrogen adatom should include a transition via Bridge and HCP2 sites with an approximately equal adsorption energy, which gives the estimate of the activation energy of nitrogen adatom diffusion equal to 1.45 eV.

TABLE II. Adsorption energies of Al, Ga, and N atoms at different adsorption sites on the pure AlN(0001) surface.

| Site \ Adatom | Al | Ga | N |
|---|---|---|---|
| FCC | 3.41 Ev | 3.13 eV | 2.36 eV |
| HCP1 | 2.58 eV | 2.53 eV | -1.53 eV |
| HCP2 | 4.06 eV | 3.75 eV | 0.93 eV |
| Bridge | 3.55 eV | 3.28 eV | 0.93 eV |

*Ab initio thermodynamic model of surface structures on the AlN(0001) surface*

To estimate the thermodynamic stability of different surface configurations, we used free Gibbs energies for calculations of system energy for the given chemical potentials $\mu_i$ of species:

$$\Omega(\mu_i, T) = E - TS - \sum_i \mu_i n_i,$$

where $E$ is the internal energy of the system, $T$ and $S$ are temperature and entropy of the system, and $n_i$ is concentration of the $i$-th species. The contribution of entropy is supposed to be equal for all condensed phases, and is taken into account only for gaseous species.

For a ternary system Al-N-H under the condition of an equilibrium with the AlN phase, the relationship between the chemical potentials of aluminum $\mu_{Al}$ and nitrogen $\mu_N$ is as follows:

$$\mu_{Al} + \mu_N = \mu_{AlN} = \mu_{Al}^{bulk} + \mu_{N2}^0 + \Delta H_f(AlN),$$

where $\Delta H_f(AlN)$ is the formation energy of AlN. Therefore, for the analysis of thermodynamic stability one can use chemical potentials of Al и H species as independent variables. The results of calculations of the phase diagram of stability of different configurations on the pure AlN(0001) surface as a function of chemical potentials of Al and H are shown in Fig. 8. The calculated phase diagram is in good agreement with those presented in [26, 27].

Based on the calculated phase diagram of stable structures on the AlN(0001) surface as a function of chemical potentials of Al and H species, one can determine stable surface structures under experimental MBE conditions. For this goal, one should express a change in chemical potentials $\Delta\mu_{Al}$ и $\Delta\mu_N$ as a function of temperature $T$ and partial pressures $P$ of species:

$$\mu_{gas} = -k_B T \ln\left[\frac{k_B T}{P}\left(\frac{2\pi m k_B T}{h^2}\right)^{3/2} \times Z_{rot} \times Z_{vib}\right],$$

where $k_B$ is the Boltzmann constant, $h$ is the Planck constant, $m$ is molecular mass, $Z_{rot}$ is rotational partition sum, and $Z_{vib}$ is vibrational partition sum.

For the conditions of AlN film deposition using the MBE method, we set hydrogen pressure in the system at $10^{-3}$ Pa. The calculated phase diagram of the most stable surface configurations on the AlN(0001) surface as a function of partial pressure of Al and temperature is shown in Fig. 8. It can be seen in the figure that, for MBE deposition conditions, only two configurations, corresponding to Al and N adatoms without hydrogen species, are thermodynamically stable. Because of low pressure in the MBE process, the chemical potential of hydrogen is about 1 eV lower for MBE deposition than that for the MOCVD process [28]. As a result, for typical MBE deposition conditions, $\Delta\mu_H \approx -2$ eV and, in accordance with the phase diagram in Fig. 8, configurations of adsorbed Al and N atoms with a coverage of ¼ monolayer are thermodynamically stable under these conditions, which is in agreement with the results of calculations in Fig. 8. We should stress, that the experimental deposition of an AlN film using the ammonia MBE process typically proceeds under an excess of ammonia, which corresponds to the N adatom region in Fig. 9.

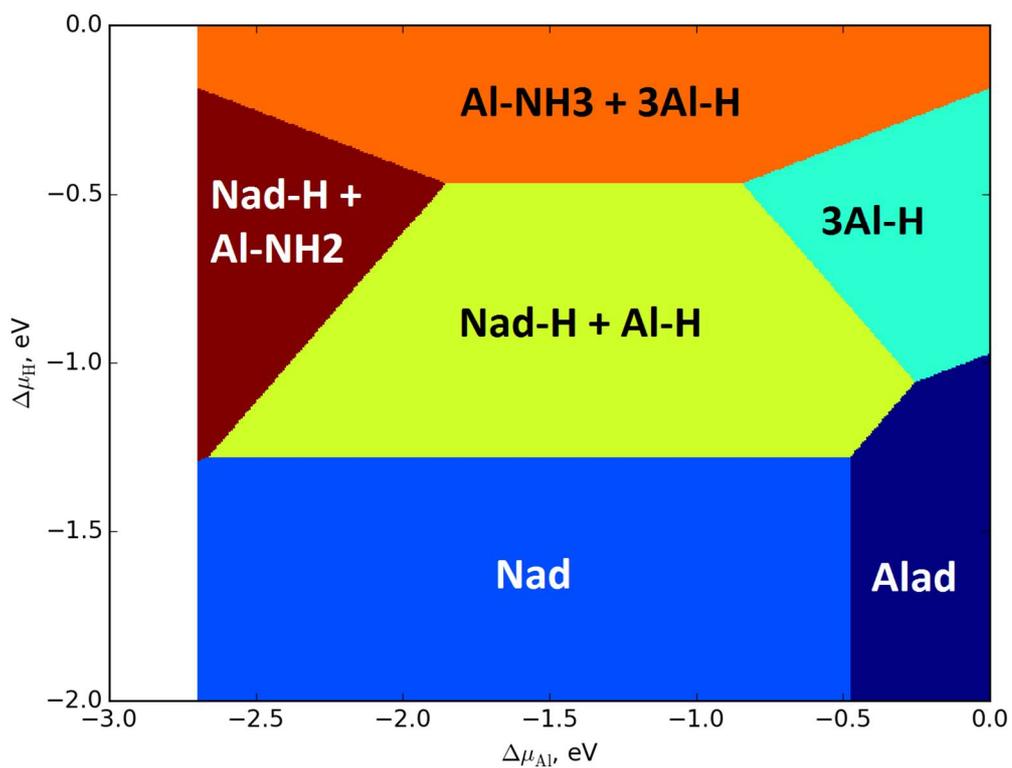

Fig. 8. Calculated phase diagram of the stability of different surface structures on the pure AlN(0001) as a function of chemical potentials of Al and hydrogen.

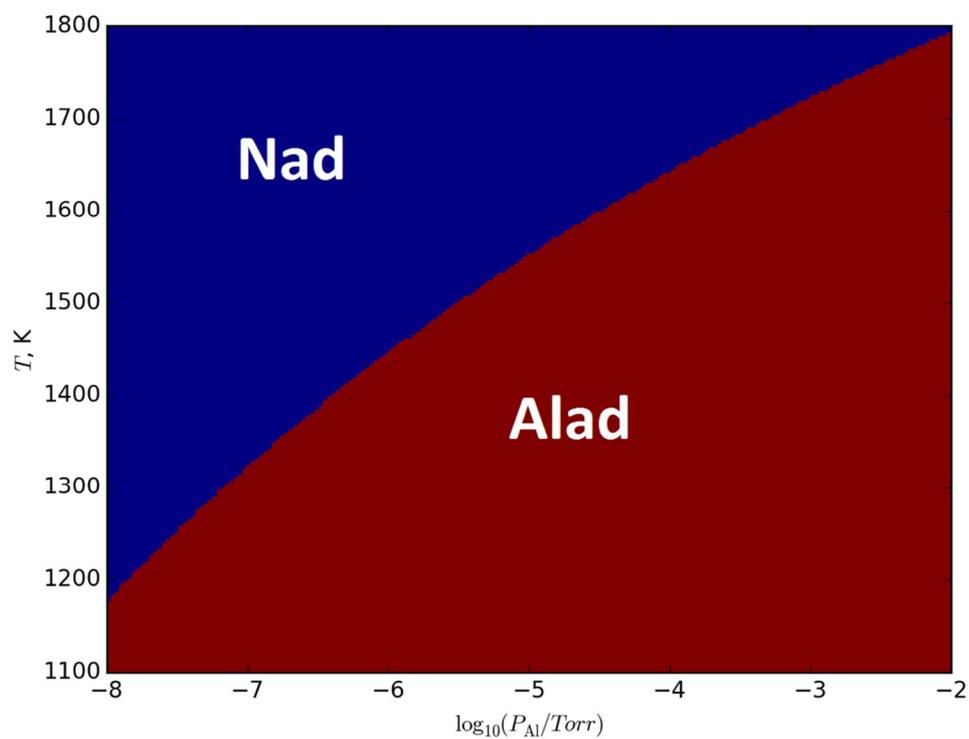

Fig 9. Calculated phase diagram of the most stable surface structures on the pure AlN(0001) surface as a function of partial pressure of Al and temperature.

## Interaction of Ga adatoms with a stepped AlN surface

For the analysis of the interaction of Ga adatoms with steps on the AlN(0001) surface, we considered a model of a stepped AlN(0001) surface, shown in Fig 7(b). The model structure contains a ribbon of the next AlN layer with a kink on the pure AlN(0001) surface. At this surface, we considered a Ga adatom on the HCP2 site co-adsorbed with one fourth of nitrogen monolayer on the FCC site, as is shown in Fig. 10(a).

The Ga adatom can attach to the kink at the step on this surface, while the nitrogen atom can then attach to Ga at the step, as is shown in Fig. 10(b). This process is exothermic and energy change in this process is –0.32 eV. This result indicates that Ga has selectivity with respect to morphological defects on the AlN(0001) surface.

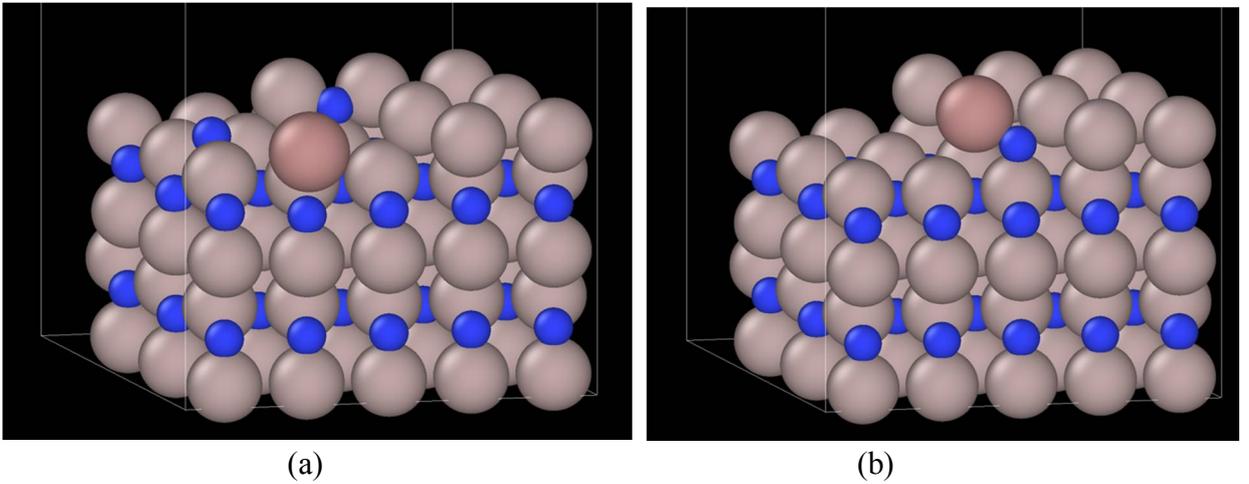

(a)            (b)

Fig. 10. Model structures for the analysis of the interaction of a Ga atom with a stepped AlN(0001) surface: (a) co-adsorption of Ga and N atoms at the terrace, (b) attachment of Ga and N atoms to the kink at the step. Notation as the same in Fig. 7.

## V. General discussion of the results.

The growth of AlGaN under the conditions of Ga desorption led to positive changes in film morphology compared to AlN growth at same growth parameters without gallium. Gallium stimulated 2D growth and allowed us to obtain films with wide smooth terraces. AlGaN films did not contain threading pores, unlike AlN. This result is interesting in contrast to GaN growth in NH3 MBE, when the evaporation of Ga leads to surface roughening [16].

Both experimental results and the data of first-principles calculations show that the binding energy of Ga depends on local surface morphology. It was found experimentally that the AlGaN phase with high Ga content formed in the vicinity of 3D surface features. Gallium atoms were accumulated by the sidewalls of 3D islands in nucleation layers and by high surface steps at subsequent growth stages. The formation of such areas shows gallium atoms preferably occupy surface sites near surface defects. Therefore, these sites provide more stable chemical bonds compared to surface sites at flat terraces. First-principles calculations also showed that the adsorption sites at steps are energetically preferable for gallium atoms compared to terrace cites. According to the calculations, the difference between the binding energies of a gallium atom at the terrace and at step is of about 0.32 eV.

It is important that gallium atoms are retained at boundaries of 3D surface features like islands and terrace steps. Under the conditions of the significant thermal desorption of gallium, the difference in binding energies favors the accumulation of additional matter at boundaries of 3D surface objects, resulting in the selective increase of the lateral growth component.

The reported mechanism has several interesting features. The first is that the accumulation of gallium in the vicinity of surface defects allows the visualization of the evolution of such defects by TEM because of a remarkable contrast produced by gallium in dark field TEM images. This allowed us to observe the movement of high terrace steps, as they left bright traces in the film bulk. The 3D–2D transition also became visible in TEM images. The other feature of the mechanism is its self-regulation, which comes from its dependence on surface morphology and from its ability to smooth out the surface. One can see in figures 3 and 4 that the accumulation of gallium was enhanced at the 3D stage and went down as soon as a smoother surface formed.

We suppose that the observed effect of the selective enhancement of lateral growth through the formation of a Ga-enriched AlGaN phase is related to the reported surfactant effect in ammonia-based epitaxial methods. The authors of [5, 7, 8] identified the surfactant effect on the basis of the following evidences: improvement of film morphology because of additives (In or Ga), stimulation of 2D growth and low average content of additives within the films. We showed that the positive effect of gallium on film morphology was observed at least up to a growth temperature of $1100°C$. At this temperature, the average mole fraction of GaN in AlGaN was only 2%, while a faster 3D–2D transition and wide smooth terraces were still observed.

Equations (5) and (6) allowed us to estimate the amount of lateral growth enhancement for the growth of AlGaN at higher temperatures with lower average concentrations of Ga. According to our consideration, the enhancement of lateral growth is due to the formation of a Ga-enriched AlGaN phase at the lateral boundaries of 3D surface features. Thus, we use Eq. (5) to calculate growth temperatures corresponding to a specific average amount of Ga. For calculation, we chose mole fractions of GaN equal to 1% and 0.1%, which are typical for the reported amount of the surfactant [5, 7, 8]. In calculations we used Al and Ga fluxes of 200 mn/h and 270 nm/h, respectively. Then the value of binding energy $E$ increased by 0.33 eV was substituted into Eq (5). For average mole fractions of GaN equal to 1% and 0.1%, we obtained that the mole fractions of GaN in the Ga-enriched AlGaN phase will be 51% and 47%, respectively. According to Eq. (6), such amount of Ga will enhance lateral growth twofold. Therefore, smoothing effect due to the selective incorporation of gallium is still significant. The Ga-enriched AlGaN phase has a small total volume, as it is restricted in the nearest neighborhood of surface defects and in the 3D nucleation layer. Therefore, the total amount of gallium measured by conventional methods, such as XRD or secondary ion mass spectrometry, would be low for films several times thicker than the nucleation layer. This estimation shows that the effects attributed to surfactants may also arise from the selective accumulation of additives. In both cases, 2D growth will be stimulated in the resulting films and the resulting content of adatoms in the films will be low.

Such a mechanism overcomes the contradiction that appears if the action of gallium or indium on growth kinetics in ammonia-based growth methods is explained by the surfactant effect only. To affect adatom diffusion barriers dramatically, the surfactant should cover the majority of surface sites. This is a conventional mode of PAMBE growth under metal-rich conditions. However, the N-rich growth excludes the formation of dense surface coverage by metal atoms. The formation of a Ga-enriched AlGaN phase does not require high coverage density by metal atoms. Therefore, N-rich conditions will not prevent the enhancement of lateral growth in contrast to the surfactant effect. This consideration does not exclude the effect of change of surface diffusion barriers due to the surfactant action of gallium and indium in MOCVD and NH3 MBE, but shows that other mechanisms should also be taken into account.

**VI. Conclusions**

In this work, we performed an experimental and theoretical study of the growth kinetics of epitaxial AlGaN films in a temperature range between full evaporation and complete incorporation of gallium.

We compared AlGaN and AlN samples grown under identical conditions by NH3 MBE. The use of gallium induced noticeable changes to film evolution, structure, and surface morphology. The formation of an AlGaN phase with higher Ga content in the areas of developed surface morphology was detected by TEM equipped with an EDXS module.

The growth rate of AlGaN at different growth temperatures was measured experimentally. Then a kinetic model was developed to analytically describe the dependence of AlGaN growth on the growth temperature. As the model was in good agreement with the experimental results, we used it to estimate the difference in the binding energies of gallium atoms at different surface cites using the data on the distribution of Ga content. First-principles analysis was also used to provide additional information on the configurations of atoms and their stability on the surface of epitaxial films.

Both experimental results and first-principles analysis showed that gallium atoms are more energetically stable at steps because of a difference in binding energies at terrace and at step adsorption sites. Both approaches gave similar value of 0.3 eV for the difference in the binding energies of Ga atoms. Under the conditions of thermal evaporation of Ga, gallium atoms are selectively retained by the boundaries of 3D surface features. The selective accumulation of gallium results in the formation of a AlGaN phase with a higher Ga content and favors an increase in the lateral growth component. The enhancement of lateral growth has a crucial effect on film morphology at the initial 3D stage of growth, promoting a faster transition to 2D growth.

We suppose that the surfactant effect in ammonia-based growth methods reported in the literature is related to the selective enhancement of lateral growth, as both mechanisms provide identical observable effects: the improvement of film morphology, stimulation of 2D growth, and a low average content of additives.

We think that our results can have several applications. The developed kinetic model for AlGaN growth rate can be used for estimating the amount of additive species and for the control of ternary film composition via temperature change. Also our results show that phase separation can be useful for the improvement of film quality at certain stages of growth, as it promotes a faster transition to 2D growth at the initial stages of AlGaN deposition, but when 2D growth is settled down, phase separation stops and gallium content drops down. Gallium accumulation at surface defects may also be a tool for the visualization of the morphology evolution by TEM, as gallium yields noticeable contrast in dark field TEM images. In our work we could observe a 3D–2D transition at the initial stage of growth and saw the movement of high surface steps during the film growth.

**Acknowledgements**

This work was partially supported by RFBR, research projects No. 15-29-01291 офи_м, and No. 16-032-00177 мол_а. The TEM analysis was done on the equipment of the Resource Center of Probe and Electron Microscopy and the Resource center of laboratory X-RAY methods (Kurchatov Complex of NBICS-Technologies, NRC "Kurchatov Institute").

**References**

[1] Stephen W. Kaun, Man Hoi Wong, Umesh K. Mishra, James S Speck. Semicond. Sci. Technol. **28**, 074001 (2013).

[2] O Ambacher. Growth and applications of Group III-nitrides. // J. Phys. D: Appl. Phys. 1998, V. 31, pp. 2653–2710

[3] Rudiger Quay, "Gallium Nitride Electronics", Springer, 2008

[4] Properties of Group III Nitrides. Ed.: J.H. Edgar. INSPEC, the Institution of Electrical Engineers, London, United Kingdom, 1994 , 1-295 p

[5] Dongjin Won, Xiaojun Weng, and Joan M. Redwing, Appl. Phys. Lett. 100, 021913 (2012).

[6] Jorg Neugebauer, Tosja K. Zywietz, Matthias Scheffler, John E. Northrup, Huajie Chen and R. M. Feenstra, Phys. Rev. Lett. 90, 056101 (2003).

[7] Stephen W Kaun, Baishakhi Mazumder, Micha N Fireman, Erin C H Kyle, Umesh K Mishra and James S Speck, Semicond. Sci. Technol. 30, 055010 (2015)

[8] T. M. Altahtamouni, J. Li, J.Y. Lin and H.X. Jiang. J. Phys. D: Appl. Phys., 45, 285103 (2012)

[9] Zahl P., Kury P. Horn-von Hoegen M., Interplay of surface morphology, strain relief, and surface stress during surfactant mediated epitaxy of Ge on Si // Appl. Phys. A., Vol. 69, N. 5, 1999, pp. 481–488

[10] S.Yu. Karpov, R.A. Talalaev, Yu. N. Makarov, N. Grandjean, J. Massies, B. Damilano, Surface Science 450, 191 (2000).

[11] S. Keller, G. Parish, P. T. Fini, S. Heikman, C.-H. Chen, N. Zhang, S. P. DenBaars, U. K. Mishra, and Y.-F. Wu. Metalorganic chemical vapor deposition of high mobility AlGaN/GaN heterostructures. Journal of Applied Physics 86, 5850 (1999)]

[12] Martin Dauelsberg, Daniel Brien, Hendrik Rauf, Fabian Reiher, Johannes Baumgartl, Oliver Häberlen, Alexander S. Segal, Anna V. Lobanova, Eugene V. Yakovlev, Roman A. Talalaev. On mechanisms governing AlN and AlGaN growth rate and composition in large substrate size planetary MOVPE reactors. Journal of Crystal Growth 393 (2014) 103–107.

[13] C. Skierbiszewski, H. Turski, G. Muziol, M. Siekacz, M. Sawicka, G. Cywinski, Z. R. Wasilewski and S. Porowski. Nitride-based laser diodes grown by plasma-assisted molecular beam epitaxy J. Phys. D: Appl. Phys. 47 (2014) 073001

[14] M. Benaissa, L. Gu, M. Korytov, T. Huault, P. A. van Aken et al. Phase separation in GaN/AlGaN quantum dots. Appl. Phys. Lett. 95, 141901 (2009)

[15] Lateral phase separation Sun et al. Appl. Phys. Lett. 87, 121914 2005,

[16] Fernández-Garrido, G. Koblmüller, E. Calleja, and J. S. Speck,. J. Appl. Phys. 104, 033541 (2008).

[17] A.R. Smith, d R. M. Feenstra, D. W. Greve, M.-S. Shin, M. Skowronski, J. Neugebauer, J. E. Northrup, App. Phys. Lett. 72, 2114 (1998).

[18] M. Schuster, P.O. Gervais, B. Jobst, W. Hosler, R. Averbeck, H. Riechert, A. Iberl and R. Stommer, J. Phys. D: Appl. Phys. 32, A56 (1999).

[19] Jarrett A. Moyer , Ran Gao , Peter Schiffer & Lane W. Martin, Scientific Reports 5, 10363 (2015).

[20] Yoshikawa, A., Xu, K., Taniyasu, Y. and Takahashi, K. (2002), Phys. Stat. Sol. (A) 190, 33 (2002).

[21] M.G.Lagally, Z.Zhang, Nature 417, 907 (2002).

[22] W. K. Burton, N. Cabrera, F.C. Frank, Phil. Trans. R. Soc. London A 243, 299 (1951).

[23] G. Kresse and J. Hafner, Phys. Rev. B 47 , 558 (1993); ibid. 49 , 14 251 (1994).

[24] G. Kresse and J. Furthmuller, Comput. Mat. Sci. 6 , 15 (1996)

[25] G. Kresse and D. Joubert, Phys. Rev. 59 , 1758 (1999).

[26] T. Akiyama, D. Obara, K. Nakamura, and T. Ito, Reconstructions on AlN Polar Surfaces under Hydrogen Rich Conditions // Japanese J. Appl. Phys. 51, 018001 (2012).

[27] Y. Kangawa, T. Akiyama, T. Ito, K. Shiraishi and T. Nakayama, Surface Stability and Growth Kinetics of Compound Semiconductors: An Ab Initio-Based Approach // Materials 6, 3309 (2013).

[28] J. E. Northrup, J. Neugebauer, Strong affinity of hydrogen for the GaN(000-1) surface: Implications for molecular beam epitaxy and metalorganic chemical vapor deposition // Appl. Phys. Lett. 85, 3430 (2004)